%% file: HIN-16-013_temp.tex
\pdfoutput=1

\documentclass[11pt,twoside,a4paper,cmspaper,final,collab]{cms-tdr}
\def\svnVersion{a2f0eff}\def\svnDate{2020/07/13}\def\cmsCernNoTag{CERN-EP-2018-213}\def\cmsCernDate{\today}\def\cmsMessage{'Published in Physical Review C as \href{http://dx.doi.org/10.1103/PhysRevC.101.064906}{\doi{10.1103/PhysRevC.101.064906}.}'}

\begin{document}\cmsNoteHeader{HIN-16-013}

\hyphenation{had-ron-i-za-tion}
\hyphenation{cal-or-i-me-ter}
\hyphenation{de-vices}
\RCS$HeadURL$
\RCS$Id$

\newlength\cmsFigWidth
\ifthenelse{\boolean{cms@external}}{\setlength\cmsFigWidth{0.98\columnwidth}}{\setlength\cmsFigWidth{0.8\textwidth}}
\ifthenelse{\boolean{cms@external}}{\providecommand{\cmsLeft}{top\xspace}}{\providecommand{\cmsLeft}{left\xspace}}
\ifthenelse{\boolean{cms@external}}{\providecommand{\cmsRight}{bottom\xspace}}{\providecommand{\cmsRight}{right\xspace}}
\newlength\cmsTabSkip\setlength{\cmsTabSkip}{1ex}

\newcommand{\PgXmXpnew}{\PgXm+ \PagXp}
\newcommand{\PgOmOpnew}{\PgOm+ \PagOp}
\newcommand{\PgLLbar}{\ensuremath{\PgL+\PagL}\xspace}
\newcommand{\pPb}{\Pp{}Pb\xspace}
\newcommand{\pp}{\Pp{}\Pp}
\newcommand{\pA}{\Pp{}A\xspace}
\newcommand{\AB}{AB\xspace}
\newcommand{\ycm}{\ensuremath{y_\mathrm{CM}}\xspace}
\newcommand{\RpPb}{\ensuremath{R_{\pPb}}\xspace}
\newcommand{\Yasym}{\ensuremath{Y_\text{asym}}\xspace}
\newcommand{\RAB}{\ensuremath{R_{\AB}}\xspace}
\newcommand{\Ncoll}{\ensuremath{N_\text{coll}}\xspace}
\newcommand{\NcollAvg}{\ensuremath{\langle\Ncoll\rangle}\xspace}
\newcommand{\EPOSLHC} {{\textsc{epos lhc}}\xspace}
\newcommand{\Vzero}{\ensuremath{\mathrm{V}^{0}}\xspace}
\newcommand{\sqrts}{\ensuremath{\sqrt{\smash[b]{s}}}\xspace}
\newcommand{\TpPb}{\ensuremath{\mathrm{T}_\mathrm{\pPb}}\xspace}
\newcommand{\TAB}{\ensuremath{\mathrm{T}_\mathrm{\AB}}\xspace}
\newcommand{\TABAvg}{\ensuremath{\langle\TAB\rangle}\xspace}

\cmsNoteHeader{HIN-16-013}
\title{Strange hadron production in \texorpdfstring{\pp and \pPb collisions at $\sqrtsNN=5.02\TeV$}{pp and pPb collisions at sqrt(s) = 5.02 TeV}}
\date{\today}

\abstract{
The transverse momentum (\pt) distributions of \PgL, \PgXm, and \PgOm\ baryons,
their antiparticles, and \PKzS\ mesons are measured in proton-proton (\pp) and
proton-lead (\pPb) collisions at a nucleon-nucleon center-of-mass energy of
5.02\TeV over a broad rapidity range. The data, corresponding to
integrated luminosities of 40.2\nbinv and 15.6\mubinv for \pp and \pPb collisions,
respectively, were collected by the CMS experiment.
The nuclear modification factor \RpPb, which is defined as the ratio of the
particle yield in \pPb collisions and a scaled \pp reference, is
measured for each particle.
A strong dependence on particle species is observed in the \pt range
from 2 to 7\GeV, where \RpPb\ for \PKzS\ is consistent with
unity, while an enhancement ordered by strangeness content
and/or particle mass is observed for the three baryons. In \pPb
collisions, the strange hadron production is asymmetric about the
nucleon-nucleon center-of-mass rapidity. Enhancements, which depend on the particle
type, are observed in the direction of the Pb beam.
The results are compared with predictions from \EPOSLHC,
which includes parametrized radial flow. The model is in
qualitative agreement with the \RpPb\ data, but fails to describe
the dependence on particle species in the yield asymmetries measured
away from midrapidity in \pPb collisions. }

\hypersetup{
pdfauthor={CMS Collaboration},
pdftitle={Strange hadron production in pp and pPb collisions at sqrt(s[NN]) = 5.02 TeV},
pdfsubject={CMS},
pdfkeywords={CMS, physics, heavy ion, spectra, strange particles}}

\maketitle

\section{Introduction}
\label{sec:intro}

The transverse momentum (\pt) distributions of the particles produced in high-energy nuclear collisions can provide
insights into the nature of the produced hot and dense matter, known as the quark-gluon plasma (QGP), and its dynamical
evolution.  Comparisons of the \pt spectra of hadrons produced in proton-proton (\pp), proton-nucleus (\pA), and
nucleus-nucleus (\AB) collisions are often used to elucidate the QGP properties. The many physical processes that
contribute to hadron production involve distinct energy scales, and therefore dominate different ranges in the \pt
distributions in various collision systems. In heavy-ion collisions, hadrons with $\pt\lesssim2\GeV$ typically reflect the properties
of the bulk system, such as the temperature at freeze-out, hadro-chemical composition,  and collective expansion velocity.
Measurements of identified hadrons at low \pt can be used to extract these
properties~\cite{PhysRevLett.78.2080, PhysRevLett.82.2471, Adler:2001aq, Adcox:2003nr, Adler:2003cb,Abelev:2013vea}.

At high \pt (${\gtrsim}8\GeV$), particles are primarily  produced through fragmentation of partons that have participated in a
hard scattering involving a large momentum transfer.
In \AB collisions that create a QGP, these partons might lose energy traversing the medium,
which would result in suppression of high-\pt hadron production. The suppression is quantified by the nuclear modification
factor, \RAB, defined as the ratio of particle yields in \AB collisions to those in \pp\ collisions, scaled
by the average number of binary nucleon-nucleon collisions, \NcollAvg, in the \AB collisions:
\begin{linenomath}
\begin{equation}
\RAB(\pt) = \frac{\rd N^\mathrm{AB}/\rd\pt}{\NcollAvg \rd N^{\pp}/\rd\pt}=\frac{\rd N^\mathrm{AB}/\rd\pt}{\TABAvg \rd\sigma^{\pp}/\rd\pt}.
\label{eqn:def_rpa}
\end{equation}
\end{linenomath}

The ratio of \NcollAvg with the total inelastic \pp\ cross section $\sigma^{\pp}$, defined  as $\TABAvg = \NcollAvg / \sigma^{\pp}$,
is known as the nuclear overlap function. Both \NcollAvg and \TABAvg can be calculated from a Glauber model of the nuclear
collision geometry~\cite{Miller:2007ri}.

In the intermediate \pt region ($2\lesssim\pt\lesssim 8\GeV$), the dominant particle production mechanism switches
from soft processes to hard scattering. For a given particle species, this
transition may happen in a momentum range that depends on the mass of the particle and on its quark composition.
Particles of greater mass are boosted to larger transverse momentum because of radial flow (common velocity
field for all particles)~\cite{Schnedermann:1993ws}, and baryon production may be enhanced ($\RAB> 1$) as a
result of hadronization by recombination~\cite{Hwa:2003bn,Fries:2003vb, Greco:2003xt}. In addition, there are several
initial-state effects that can result in $\RAB \neq 1$. Momentum broadening from multiple scattering of projectile
partons by the target nucleus before undergoing a hard scattering~\cite{Wang:2003vy,Kang:2012kc} can cause an enhancement.
Alternatively, nuclear shadowing~\cite{NuclEff:1994}, \ie,
suppression of the parton distribution functions in the nucleus relative to those in the proton in the small parton fractional
momentum range ($x < 0.01$), can lead to
suppression in hadron production.
The study of nuclear modification factors over a broad
momentum range and for multiple particle species is a valuable
tool for disentangling different effects and for constraining theoretical models.

Traditionally, \pA and deuteron-nucleus (dA) collisions have been considered
as reference systems that do not produce a hot
QCD medium~\cite{Adler:2003ii,Back:2003ns, Adams:2003im, Arsene:2003yk}, and
therefore would only carry information about cold nuclear matter initial-state effects. However,
in the last few years there have been extensive studies of two- and multiparticle azimuthal correlations in high-multiplicity \pp and
\pPb\ collisions at the LHC~\cite{CMS:2012qk, Chatrchyan:2013nka,Khachatryan:2015waa, Khachatryan:2016txc}, which indicate collective
behavior similar to that observed in heavy-ion collisions, where it 
is attributed to collective flow in the QGP. Recent measurements 
from the BNL Relativistic Heavy Ion Collider (RHIC) use high-multiplicity {\Pp}Au~\cite{Aidala:2016vgl}, dAu~\cite{Adare:2014keg}, and $^3$HeAu collisions~\cite{Adare:2015ctn} to study the effects of
the initial geometry on the final-state particle correlations. They find that hydrodynamic models that include short-lived QGP droplets
provide simultaneous quantitative description of the measurements~\cite{PHENIX:2018lia}. Additionally, measurements of strange-particle production by the ALICE
Collaboration~\cite{Adam:2015vsf, ALICE:2017jyt} indicate strangeness enhancement in \pPb\ and high-multiplicity \pp\ collisions---a
signature that has long been considered an important indication of QGP formation~\cite{Rafelski:1982pu}. Measurements of low-\pt spectra
of strange particles produced in high multiplicity small-system collisions~\cite{Khachatryan:2016yru, Adam:2015vsf} are consistent with the
presence of radial flow ~\cite{Chatrchyan:2013eya}. On the other hand,
jet quenching is not observed at high \pt in \pPb\ collisions~\cite{Khachatryan:2016xdg, Adam:2015hoa,Abelev:2014dsa, Aad:2016zif, Khachatryan:2015xaa}.
Thus, further studies of the rapidity and \pt dependence of strange-particle
production from low to high \pt can provide significant information
on the nature of the QCD medium produced in small systems.

In \pPb collisions, radial flow,  nuclear shadowing, and multiple scattering are all expected to have different effects on particle production in the
forward (\Pp-going) and backward (Pb-going) rapidity regions. Radial flow is expected to be greater in the Pb-going than the
\Pp-going direction and therefore to produce a stronger mass dependence on the Pb-going side~\cite{Bozek:2013sda, Pierog:2013ria}.
The effect of nuclear shadowing is expected to be more prominent in the \Pp-going  direction, where smaller $x$ fractions are accessed
in the nucleus. This should result in larger \RpPb values in the Pb-going
as compared with the \Pp-going direction.

The effect of parton multiple scattering is
not completely understood, and has been shown to depend on multiple factors, \eg, whether the scatterings are elastic, inelastic, coherent or
incoherent~\cite{Wang:2003vy,Kang:2013ufa}. These predictions can be tested with measurements of \RpPb in the
\Pp- and Pb-going directions separately, and of the particle yield rapidity asymmetry \Yasym in \pPb\ collisions, where
\begin{linenomath}
\begin{equation}
\Yasym(\pt) =
\frac{\rd^{2}N(\pt)/\rd \ycm \rd\pt|_{\ycm \in [-b, -a]}}{\rd^{2}N(\pt)/\rd \ycm \rd\pt|_{\ycm \in [a, b]}}.
\label{eqn:def_yasym}
\end{equation}
\end{linenomath}
Here,  \ycm is the rapidity computed in the center-of-mass frame of the colliding nucleons, $a$ and $b$ are always non-negative and, by definition, refer to the proton beam direction.

This paper presents measurements of strange hadron \pt\ spectra at $\abs{\ycm}<1.8$, $-1.8<\ycm<0$, and $0<\ycm<1.8$ in \pp\ and \pPb\ collisions at $\sqrtsNN=5.02\TeV$.
These measurements are shown for the \PKzS\ and the sum of \PgLLbar,
\PgXmXpnew, and \PgOmOpnew (hereafter referred to as \PgL, \PgXm, and \PgOm, respectively).
Based on these spectra, \RpPb for each particle species is studied as a function of \pt in the three rapidity ranges above.
Because of limitations in the size of the data sample, the \RpPb of the \PgOm\ baryon is studied  in the
range $\abs{\ycm}<1.8$. To study the rapidity dependence in strange hadron production in \pPb collisions, the \PKzS\ and \PgL\
spectra are measured in several additional rapidity ranges.
The \Yasym is evaluated  for $0.3<\abs{\ycm}<0.8$, $0.8<\abs{\ycm}<1.3$,
and $1.3<\abs{\ycm}<1.8$.
The results are compared with predictions from
the \EPOSLHC model, which includes collective flow in \pp\ and \pPb collisions.

\section{The Compact Muon Solenoid detector}
\label{sec:cmsdetector}

The central feature of the Compact Muon Solenoid (CMS) apparatus is a superconducting solenoid of 6\unit{m} internal diameter,
providing a magnetic field of 3.8\unit{T}. Within the solenoid volume are a silicon pixel and strip
tracker, a lead tungstate crystal electromagnetic calorimeter, and a brass and scintillator
hadron calorimeter (HCAL), each composed of a barrel and two endcap sections. Forward calorimeters
extend the pseudorapidity ($\eta$) coverage provided by the barrel and endcap detectors.
The silicon tracker measures charged particles within the range $\abs{\eta} < 2.5$.
It consists of 1440 silicon pixel and 15\,148 silicon strip detector modules.
The pixel detector comprises three barrel layers and two forward disks on each side
of the interaction point.
For nonisolated
particles of $1 < \pt < 10\GeV$ and $\abs{\eta} < 1.4$, the track resolutions are
typically 1.5\% in \pt and 25--90 (45--150)\mum in the transverse (longitudinal) impact parameter~\cite{TRK-11-001}.
The forward hadron (HF) calorimeter uses steel as an absorber and quartz fibers as the sensitive
material. The two halves of the HF are located 11.2\unit{m} from the interaction region, one on
each end, and together they provide coverage in the range $3.0 < \abs{\eta} < 5.2$.
A more detailed description of the CMS detector, together with a definition of the coordinate system
used and the relevant kinematic variables, can be found in Ref.~\cite{Chatrchyan:2008zzk}.
The Monte Carlo (MC) simulation of the particle propagation and detector response is
based on the \GEANTfour~\cite{Agostinelli:2002hh} program.

\section{Data samples and event selection}
\label{sec:evtsel}

Minimum bias (MB) \pp\ and \pPb data used in this analysis were collected in 2015 and 2013 at $\sqrtsNN=5.02\TeV$, corresponding to integrated luminosities
of 40.2\nbinv and 15.6\mubinv, respectively. In \pPb collisions,
the beam energies were 4\TeV for protons and 1.58\TeV per nucleon
for lead nuclei. The data were collected in two different run conditions: one with the protons circulating in the clockwise direction in the LHC ring, and one with them circulating in the
counterclockwise direction. By convention, the proton beam rapidity is taken to be positive when combining the data from the two run configurations. Because of the asymmetric beam conditions, the nucleon-nucleon center-of-mass in the \pPb collisions moves with speed $\beta = 0.434$ in the laboratory frame. As a consequence, a massless particle emitted at $\ycm = 0$ will be detected at a rapidity of $0.465$ in the laboratory frame.

The triggers and event selections are the same as those discussed
for \pp\ collisions in Refs.~\cite{Khachatryan:2016odn, Khachatryan:2016bia}, requiring one energy deposit above the readout threshold of 3\GeV on either side of the HF calorimeters. The MB \pPb events are triggered by requiring at least one reconstructed track
with $\pt >0.4\GeV$ in the pixel detector.

In the subsequent analysis of both collision systems, events are selected by requiring at least one reconstructed collision vertex with two or more
associated tracks. All vertices are required to be within 15\unit{cm} of the nominal interaction point along the beam axis and 0.15\unit{cm} transverse to the beam axis direction. Beam-related background is suppressed by rejecting events in which less than 25\% of all reconstructed tracks satisfy the high-purity selection defined in Ref.~\cite{TRK-11-001}. In addition, having at least one HF calorimeter tower on each side of the HF with more than 3\GeV of total energy is required for \pPb collisions to further remove background events. There is a 3\% probability to have at least one additional interaction in the same bunch crossing (pileup) in the \pPb data sample.
The procedure used to reject pileup events in \pPb collisions is
described in Ref.~\cite{Chatrchyan:2013nka}. It is based on the
number of tracks associated with each reconstructed vertex and the
distance between different vertices. The pileup-rejection efficiency
is found to be $92\%\pm2\%$, which is confirmed by using a low pileup
data sample. The average pileup (the mean of the Poisson distribution
of the number of collisions per bunch crossing) is approximately 0.9
in \pp\ collisions. Following the same procedure as in
Ref.~\cite{Khachatryan:2016odn}, all the reconstructed vertices are
selected to extract the \pp\ strange-particle spectra.
The \pp\ integrated luminosity~\cite{CMS-PAS-LUM-16-001} is used
to normalize the spectrum in \pp\ collisions.

The \PYTHIA 8.209 generator~\cite{Sjostrand:2014zea} with the underlying event tune CUETP8M1~\cite{Khachatryan:2015pea} is used to
simulate the selection efficiency in \pp\ collisions. The efficiency to identify inelastic events is 95\%. For \pPb collisions, the
selection efficiency is estimated with respect to a detector-independent class of collisions termed ``double-sided'' (DS) events, which are
very similar to those that pass the HF selection criteria described above. A DS event is defined as a collision producing at least one
particle of lifetime $c\tau > 10^{-18} \unit{m}$ with energy $E > 3\GeV$ in the region $3 < \eta < 5$, and
another such particle in the region $-5 < \eta < -3$. In a simulated sample of \pPb\ DS events produced using
version 1.383~\cite{Gyulassy:1994ew} of the \HIJING\ MC generator~\cite{Wang:1991hta}, the above selection has a 99\% selection efficiency. 
A similar study using the \EPOSLHC\ generator shows 
less than 1\% difference. In MC samples produced 
by \EPOSLHC\ and \HIJING, DS events correspond to 94\%--97\% of
the hadronic inelastic \pPb collisions. A procedure similar to that in Refs.~\cite{Khachatryan:2016odn, Khachatryan:2015xaa} is used
to correct the strange-particle spectra in \pp\ and \pPb\ collisions to spectra for inelastic collisions and DS events, respectively,
with multiplicity-dependent correction factors. The values of \RpPb will decrease by 3\%--6\% if the normalization of the \pPb spectra are
corrected for the efficiency of detecting inelastic collisions instead of DS events.

\section{Particle reconstruction and yields}
\label{sec:reco_corr}

\begin{figure*}[th!]
\centering
\includegraphics[width=0.9\textwidth]{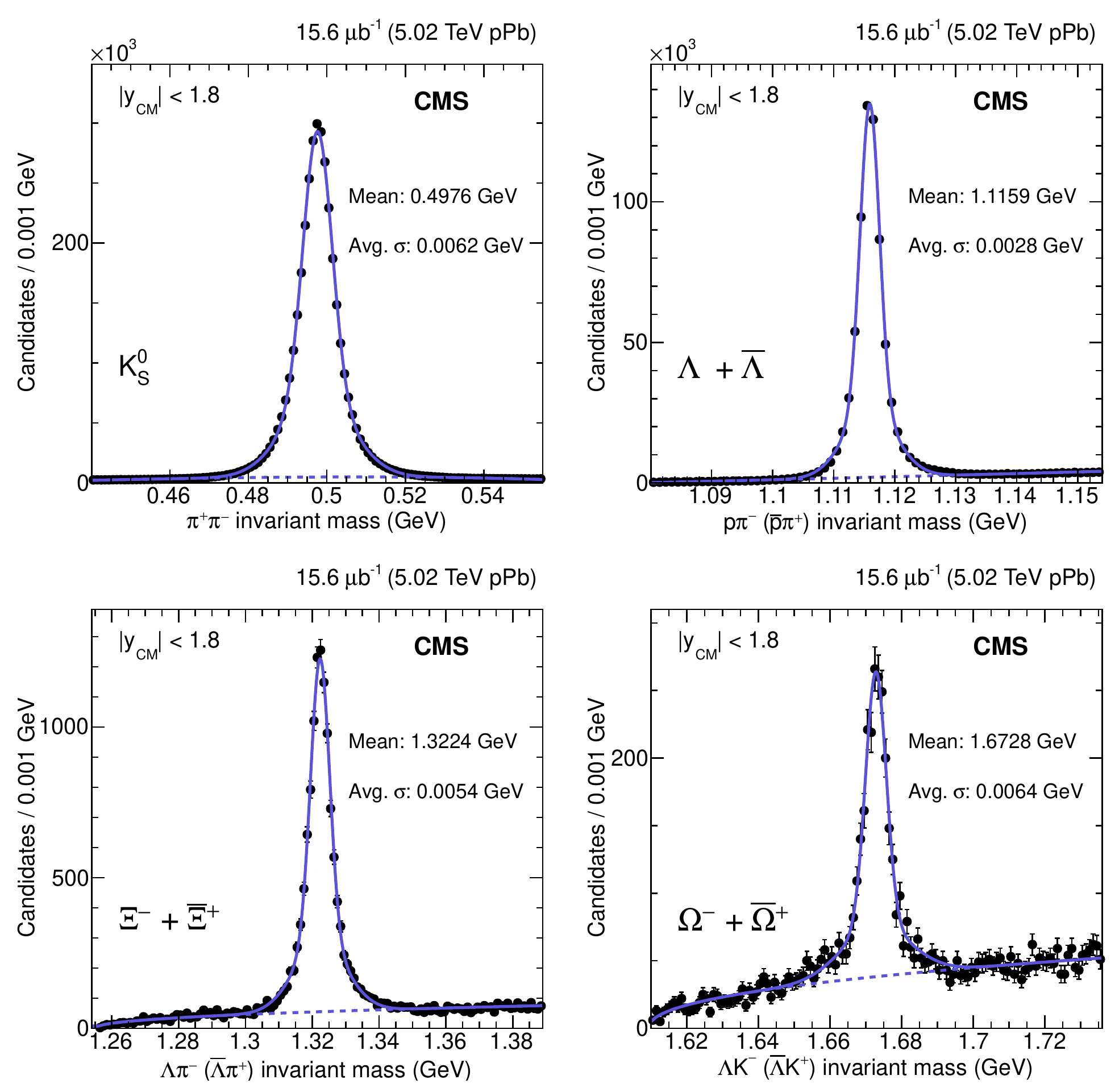}
\caption{ Invariant-mass distribution of \PKzS\ (upper left), \PgLLbar\ (upper right), \PgXmXpnew\ (lower left),
and \PgOmOpnew\ (lower right) candidates within $\abs{\ycm} < 1.8$ in \pPb collisions.
The solid lines show the results of fits described in the text. The dashed lines indicate the fitted background component.}
\label{fig:invmasspeak}
\end{figure*}

The \PKzS, \PgL, \PgXm, and \PgOm\ candidates in this paper are identified and analyzed following
the procedure used in previous analyses~\cite{Khachatryan:2011tm, Khachatryan:2016yru}. The \PKzS\ and \PgL\
(generally referred to as \Vzero) candidates are reconstructed via their decay topology by combining pairs of
oppositely charged tracks that are displaced from the primary vertex to define a secondary vertex. The mass ranges
are indicated by the horizontal axes of Fig.~\ref{fig:invmasspeak}. In the \PKzS\ reconstruction, the two tracks
are assumed to be pions. For \PgL\ reconstruction, the track with lower momentum is assumed to be a pion, while the
one with higher momentum is assumed to be a proton. To optimize the reconstruction of \Vzero particles,
requirements are applied to the three-dimensional (3D) distance of closest approach (DCA) significance of the
\Vzero decay products with respect to the primary vertex. This significance, defined as the 3D DCA between the decay
products and the primary vertex divided by its uncertainty, must be larger than two for both daughter tracks.
To further reduce the background from random combinations of tracks, the 3D DCA significance of the \Vzero candidates
with respect to the primary vertex cannot exceed 2.5. Because of the long lifetime of the \Vzero particles, the 3D decay
length significance, which is the 3D distance between the primary and \Vzero vertices divided by its uncertainty, must
be larger than three. To remove \PKzS\ candidates misidentified as \PgL\ particles, the \PgL\ candidate mass assuming both
tracks to be pions must differ from the nominal \PKzS\ mass value~\cite{Patrignani:2016xqp} by more than 20\MeV. A similar
procedure is done to remove \PgL\ candidates misidentified as \PKzS\ particles. To remove photon conversions to an
electron-positron pair, the \Vzero candidate mass must exceed 15\MeV if the tracks are both assumed to have the electron mass.

For the \PgXm\ and \PgOm\ baryon reconstruction, a previously reconstructed \PgL\ candidate is combined with an additional
charged track carrying the correct charge sign, to define a common secondary vertex. This track is assumed to be a pion (kaon)
in \PgXm\ (\PgOm) reconstruction. Since the \PgL\  candidate in the reconstruction of \PgXm\ and \PgOm is a secondary particle,
the 3D separation significance between the \PgL\ candidate vertex and the primary vertex is required to be larger than 10.
Additionally, the 3D DCA  significance requirement for the pion track from the \PgL\ candidate is increased from two to three,
and this has the effect of reducing the background in the reconstruction of \PgXm\ and \PgOm. The 3D DCA significance of a pion
(kaon) track from the \PgXm\ (\PgOm) baryon decay with respect to the primary vertex is required to be larger than four. To ensure
that the reconstructed \PgXm\ and \PgOm candidates are primary particles, their 3D DCA significance with respect to the primary vertex
is required  to be less than three.

The invariant-mass distributions of reconstructed \PKzS, \PgL, \PgXm, and \PgOm\ candidates in the range $\abs{\ycm} < 1.8$ are shown in Fig.~\ref{fig:invmasspeak} for \pPb events. Prominent mass peaks are visible, with little background. The solid lines show the results of a maximum
likelihood fit. In this fit, each strange-particle mass peak is modeled using a sum of two Gaussian functions with a common mean.
The ``average $\sigma$'' values in Fig.~\ref{fig:invmasspeak} are the square root of the weighted average of the variances of the two
Gaussian functions. The background is modeled by using a quadratic function for the \PKzS\ mesons, and with the analytic form $Cq^D$ for
the baryons to mimic the available phase-space volume, where $q$ is the difference between the mass of the mother candidate and the sum
of the assumed two daughter track masses, and C and D are free parameters. These fit functions are found to provide a reasonable
description of the signal and background with relatively few free parameters. The fits are performed over the mass ranges indicated by
the limits of the horizontal axes in each panel of Fig.~\ref{fig:invmasspeak} to obtain the raw strange-particle
yields$N^\text{raw}_{\PKzS}$, $N^\text{raw}_{\PgL}$, $N^\text{raw}_{\PgXm}$, and   $N^\text{raw}_{\PgOm}$.

The raw strange-particle yield is corrected for the branching fraction ($B$), acceptance ($\alpha$), and reconstruction
efficiency ($\epsilon$), using simulations based on the \EPOSLHC\ event generator~\cite{Pierog:2013ria} and a \GEANTfour\ model of
the CMS detector. The corrected yield, $N^{\text{corr}}_{\PKzS}$, 
$N^{\text{corr}}_{\PgL}$, $N^{\text{corr}}_{\PgXm}$, $N^{\text{corr}}_{\PgOm}$, is given by
\begin{linenomath}
\begin{equation}
\label{eq:efficiency}
\begin{split}
N^{\text{corr}}_{\PKzS} = \frac{N^{\text{raw}}_{\PKzS}}{B \, \alpha \, \epsilon},\\
N^{\text{corr}}_{\PgL} = \frac{N^{\text{raw}}_{\PgL}}{B \, \alpha \, \epsilon},\\
N^{\text{corr}}_{\PgXm} = \frac{N^{\text{raw}}_{\PgXm}}{B \, \alpha \, \epsilon},\\
N^{\text{corr}}_{\PgOm} = \frac{N^{\text{raw}}_{\PgOm}}{B \, \alpha \, \epsilon},
\end{split}
\end{equation}
\end{linenomath}
where $B \, \alpha \, \epsilon$ is obtained by the ratio of reconstructed yield to generated yield of prompt strange particles in
MC simulations. The corrections are obtained separately in each rapidity range under study.

The raw \PgL\ particle yield also contains a contribution from decays of \PgXm\ and \PgOm\ particles. This ``nonprompt'' contribution is
largely determined by the relative ratio of \PgXm\ to \PgL\ yield since the contribution from $\PgOm$ particles is negligible. While
stringent requirements on the significance of the 3D DCA for the \PgL\ candidates with respect to the primary vertex remove a large
fraction of nonprompt \PgL\ candidates, up to 4\% of the \PgL\ candidates from simulations are found to be nonprompt at intermediate \pt.
The method used to account for the nonprompt \PgL\ contribution is the same as in the previous analysis~\cite{Khachatryan:2016yru}. If the
ratio of \PgXm\ to \PgL\ yield is modeled precisely in MC generators, contamination of nonprompt \PgL\ particles will
be eliminated in the correction procedure using Eq.~\eqref{eq:efficiency}. Otherwise, an additional correction for the residual effect
is necessary. As the \PgXm\ particle yields are explicitly measured in this analysis, this residual correction factor can be derived from
data as:
\begin{linenomath}
\begin{equation}
\label{PromptEfficiency}
f^{\text{residual}}_{\PgL, \; \mathrm{np}} = 1+f^\mathrm{raw, \; MC}_{\PgL, \; \mathrm{np}} \, \left ( \frac{N^{\text{corr}}_{\PgXm}/N^{\text{corr}}_{\PgL}}{N^{\mathrm{MC}}_{\PgXm}/N^{\mathrm{MC}}_{\PgL}}-1 \right ),
\end{equation}
\end{linenomath}
where $f^\mathrm{raw, \; MC}_{\PgL, \; \mathrm{np}}$ denotes the fraction of nonprompt \PgL\ candidates in the reconstructed sample, and is obtained from MC simulation. The $N^{\text{corr}}_{\PgXm}/N^{\text{corr}}_{\PgL}$ and $N^{\mathrm{MC}}_{\PgXm}/N^{\mathrm{MC}}_{\PgL}$ terms are the \PgXm-to-\PgL\ ratios from the data after applying corrections in Eq.~\eqref{eq:efficiency}, and from generator-level MC simulations, respectively. The final measured \PgL\ particle yield is given by $N^{\text{corr}}_{\PgL}/f^{\text{residual}}_{\PgL, \; \mathrm{np}}$. Based on studies using \EPOSLHC, which has a similar \PgXm-to-\PgL\ ratio as the data, the
residual nonprompt contributions to \PgL\ yields are found to be negligible. Note that $N^{\text{corr}}_{\PgL}$ used 
in Eq.~\eqref{PromptEfficiency} is first derived by 
using Eq.~\eqref{eq:efficiency}, which in principle contains the residual nonprompt \PgL\  contributions. Therefore, by applying Eq.~\eqref{PromptEfficiency} in an iterative fashion, $N^{\text{corr}}_{\PgL}$ will approach a result corresponding to prompt \PgL\ particles. A second iteration of the correction procedure was found to have an effect of less than 0.1\% of the \PgL\ baryon yield, and hence was not pursued. The nonprompt contributions to \PgXm\ and \PgOm\ baryon yields are found to be negligible,
since the absolute yields and branching ratios of the hadrons that feed into them are much smaller than those for \PgL\  baryons.

\section{Systematic uncertainties}
\label{sec:sys}

The dominant sources of systematic uncertainty are associated with the strange-particle reconstruction,
especially the efficiency determination. Tables~\ref{tab:syst-table-summary} and~\ref{tab:syst-table-summary-yasym} summarize
the sources of systematic uncertainties in the \PKzS, \PgL, \PgXm, and \PgOm\ \pt\ spectra, \RpPb, and \Yasym for different \ycm ranges
in both \pp\ and \pPb\ collisions.

\begin{table*}[htb]
\topcaption{Summary of different sources of systematic uncertainties in \PKzS, \PgL, \PgXm, and \PgOm\ \pt\ spectra and \RpPb
measurements for different \ycm ranges in both \pp\ and \pPb collisions. The ranges quoted cover both the \pt\ and the rapidity
dependence of the uncertainties.}
\centering
\begin{scotch}{l*{4}{c}}
Source         & {\PKzS\ (\%)} & {\PgL\ (\%)} & {\PgXm\ (\%)} & {\PgOm\ (\%)}\\
\hline
Yield extraction &0--2 &0--4 &2 &3\\
Selection criteria &1--4  &1--5  &3  &6 \\
Momentum resolution &1 &1 &1 &1\\
Tracking efficiency &8 &8 &12 &12\\
Feed-down correction    &    &2--3   &    &  \\
Pileup effect (\pp\ only) &1--2.3 &1--2 &3 &3 \\
Beam direction (\pPb only) &1--4 &1--5 &3  &4\\
Integrated lum. (\pp\ only) &2.3 &2.3 &2.3 &2.3\\
$\langle \TpPb \rangle$ (for \RpPb) &4.8 &4.8 &4.8 &4.8\\[\cmsTabSkip]
Total (yields in \pp\ coll.) &8.6--9.3 &8.9--10.6 &13.1 &14.3\\
Total (yields in \pPb\ coll.)  &8.2--10.1 &8.6--12.3 &13.8 &15.1\\[\cmsTabSkip]
Total (\RpPb) &3.1--5.6 &4.3--10.4 &6.8 &10.8\\
\end{scotch}
\label{tab:syst-table-summary}
\end{table*}

\begin{table}[htb]
\topcaption{ Summary of systematic uncertainties in the \Yasym measurements in \pPb collisions. The ranges
quoted cover both the \pt\ and the rapidity dependence of the uncertainties. Because of limitations in the size of the data sample, the \Yasym of \PgXm\ and \PgOm\ are not presented.}
\centering
\begin{scotch}{l*{2}{c}}
Source       & {\PKzS\ (\%)} & {\PgL\ (\%)} \\
\hline
Yield extraction &   &0--3\\
Selection criteria &1--5 &1--6\\
Momentum resolution &1 &1 \\
Feed-down correction  &   &2--3 \\
Beam direction &2--4 &2--6\\[\cmsTabSkip]
Total (\Yasym)  &2.4--6.5 &3.2--9.3\\
\end{scotch}
\label{tab:syst-table-summary-yasym}
\end{table}

The systematic uncertainty from the yield extraction is evaluated with different background fit functions and methods for extracting the yields.
The background fit function is varied to a third-order polynomial for the systematic studies.
The yields are compared between integrating over the signal functions and counting the yield from the signal region of the histograms.
On the basis of these studies, systematic uncertainties of 0\%--4\% are assigned to the yields. Systematic effects related to the selection
of the strange-particle candidates are evaluated by varying the selection criteria, resulting in an uncertainty of 1\%--6\%. The impact of
finite momentum resolution on the spectra is estimated using the \EPOSLHC\ event generator. Specifically, the generator-level \pt\ spectra
of the strange particles are smeared by the momentum resolution, which is determined from the momentum difference between the generator-level
and the matched reconstructed-level particles. The difference between the smeared and original spectra is less than 1\%.
The systematic uncertainty in determining the efficiency of a single track is 4\%~\cite{CMS-PAS-TRK-10-002}. The tracking efficiency
is strongly correlated with the lifetime of a particle, because when and where a particle decays determine how efficiently the detector
captures its decay products. We observe agreement of the strange particle lifetime distribution ( $c\tau$) between data and simulation,
which provides a cross-check. This translates into a systematic uncertainty in the reconstruction
efficiency of 8\% for the \PKzS\ and \PgL\ particles, and 12\% for the \PgXm\ and \PgOm\ particles. The systematic uncertainty
associated with a feed-down effect for the \PgL\ candidate spectra is evaluated through propagation of the systematic uncertainty
in the $N^{\text{corr}}_{\PgXm}/N^{\text{corr}}_{\PgL}$ ratio in Eq.~\eqref{PromptEfficiency} to the $f^{\text{residual}}_{\PgL, \; \mathrm{np}}$ factor,
and is found to be 2\%--3\%. Systematic uncertainty introduced by pileup effects for \pp\ data is estimated to be 1\%--3\%. This uncertainty
is evaluated through the comparison of strange-particle spectra between data with low and high pileup. The uncertainty associated with
pileup is negligible for the \pPb data. In \pPb collisions, the direction of the \Pp and Pb beams were reversed during the course of the
data collection. A comparison of the particle \pt\ spectra in both data periods yields an uncertainty of 1\%--5\%. The uncertainty in the
integrated luminosity for \pp\ collisions is 2.3\%~\cite{CMS-PAS-LUM-16-001}. As in Ref.~\cite{Khachatryan:2015xaa}, the uncertainty in $\langle \TpPb \rangle$ is 4.8\%.

Since the same tracking algorithm is used in the \pp\ and \pPb\ data reconstruction, the uncertainties in the tracking efficiency
largely cancel in the \RpPb ratio and are negligible compared with other sources of systematic uncertainty, which are uncorrelated
between the two collision systems and are summed in quadrature. The overall uncertainty in \RpPb for the different particle species
are listed in the bottom row of Table~\ref{tab:syst-table-summary}. These numbers exclude the luminosity and $\langle \TpPb \rangle$
uncertainties, which are common to all data points. 

The uncertainties in \Yasym are evaluated in a similar way as for the particle spectra, but the effects of the different sources of
uncertainty are considered directly in the values of \Yasym. The tracking efficiency largely cancels in the ratio, while the effects
from the detector acceptance are accounted for by comparing the data sets taken with different beam directions. The remaining
uncertainties are uncorrelated and are summed up in quadrature, as detailed in Table~\ref{tab:syst-table-summary-yasym}.

\section{Results}
\subsection{Transverse momentum spectra and nuclear modification factor}

\begin{figure*}[ht!]
\centering
\includegraphics[width=0.9\textwidth]{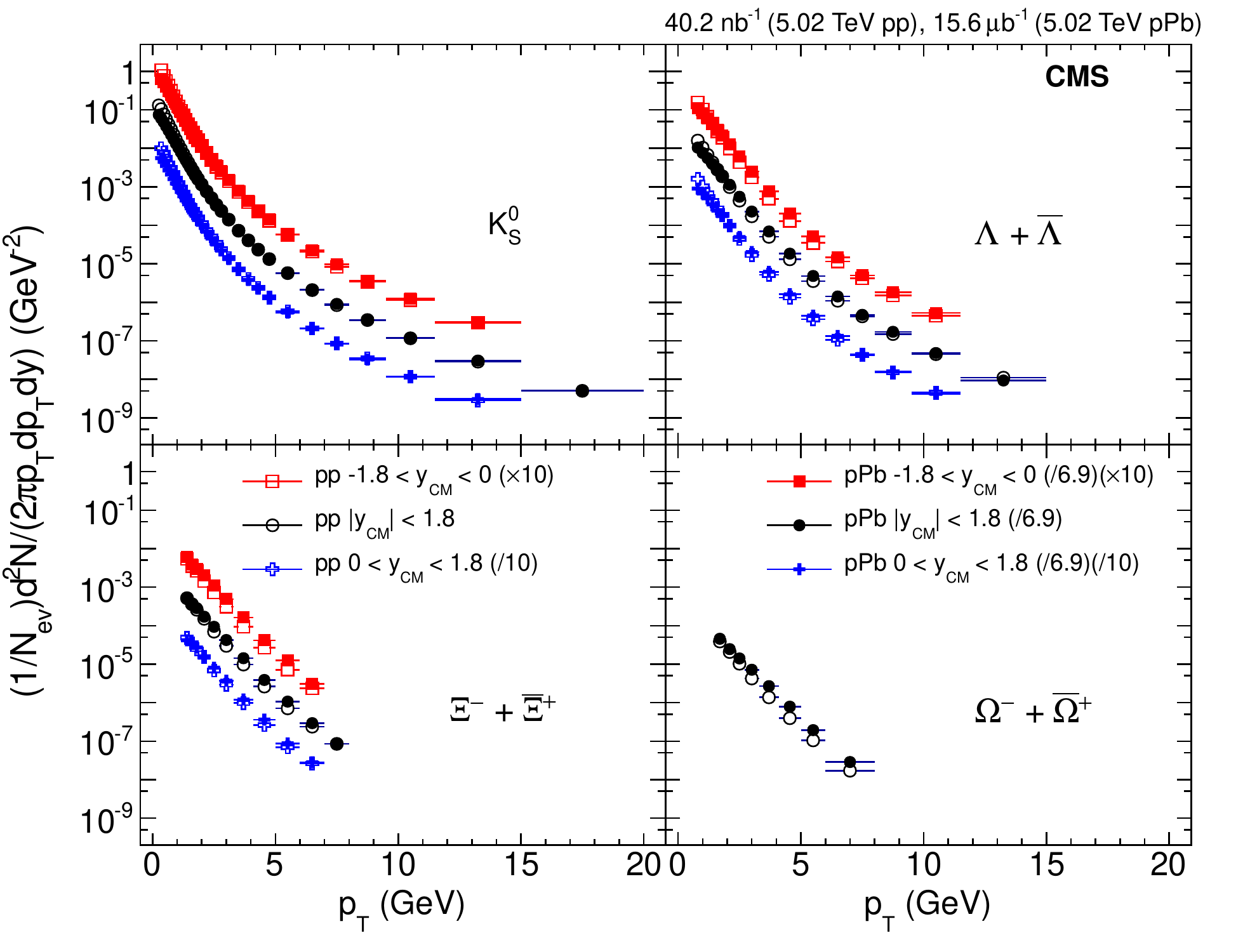}
\caption{The invariant \pt-differential spectra of \PKzS\ (upper left), \PgLLbar\ (upper right), \PgXmXpnew\ (lower left), and \PgOmOpnew\ (lower right) for $\abs{\ycm} < 1.8$, $-1.8 < \ycm < 0$, and $0 < \ycm < 1.8$ in \pp\ and \pPb collisions at $\sqrtsNN=5.02\TeV$. Spectra for different \ycm ranges are scaled by factors of powers of 10, with $\abs{\ycm} < 1.8$ not scaled.
To compare the strange-particle spectra in \pp\ and \pPb collisions directly,
the spectra in \pPb collisions are divided by 6.9, which is the average number of binary nucleon-nucleon collisions.
The vertical bars correspond to statistical uncertainties, which are usually smaller than the marker size, while the horizontal bars represent the bin width.
}
\label{fig:spectra_rpa}
\end{figure*}

The invariant \pt-differential spectra of \PKzS, \PgL, \PgXm, and \PgOm\ particles with
$\abs{\ycm} < 1.8$, $-1.8<\ycm<0$, and $0<\ycm<1.8$ in \pp\ and \pPb collisions at $\sqrtsNN=5.02\TeV$ are
presented in Fig.~\ref{fig:spectra_rpa}. For \RpPb calculations, the \pp\ spectrum is measured as a differential cross section with normalization determined from
the integrated luminosity. To convert the cross-section to a per-event yield for comparison on the same figure,
it is divided by $70 \pm 5\unit{mb}$~\cite{Patrignani:2016xqp, Khachatryan:2016odn},
which corresponds to the total inelastic \pp\ cross section.
To compare the strange-particle spectra in \pp\ and \pPb collisions directly, the spectra in \pPb collisions are divided
by the average number of binary nucleon-nucleon collisions, $\NcollAvg = 6.9 \pm 0.5$, 
which is obtained from a Glauber MC simulation~\cite{Miller:2007ri}.
The nuclear radius and skin depth utilized 
are $6.62 \pm 0.06\unit{fm}$ and $0.546 \pm 0.010\unit{fm}$, respectively, 
and a minimal distance between the nucleons 
of $0.04 \pm 0.04\unit{fm}$ is imposed~\cite{Khachatryan:2016odn}.

\begin{figure}[ht!]
\centering
\includegraphics[width=\cmsFigWidth]{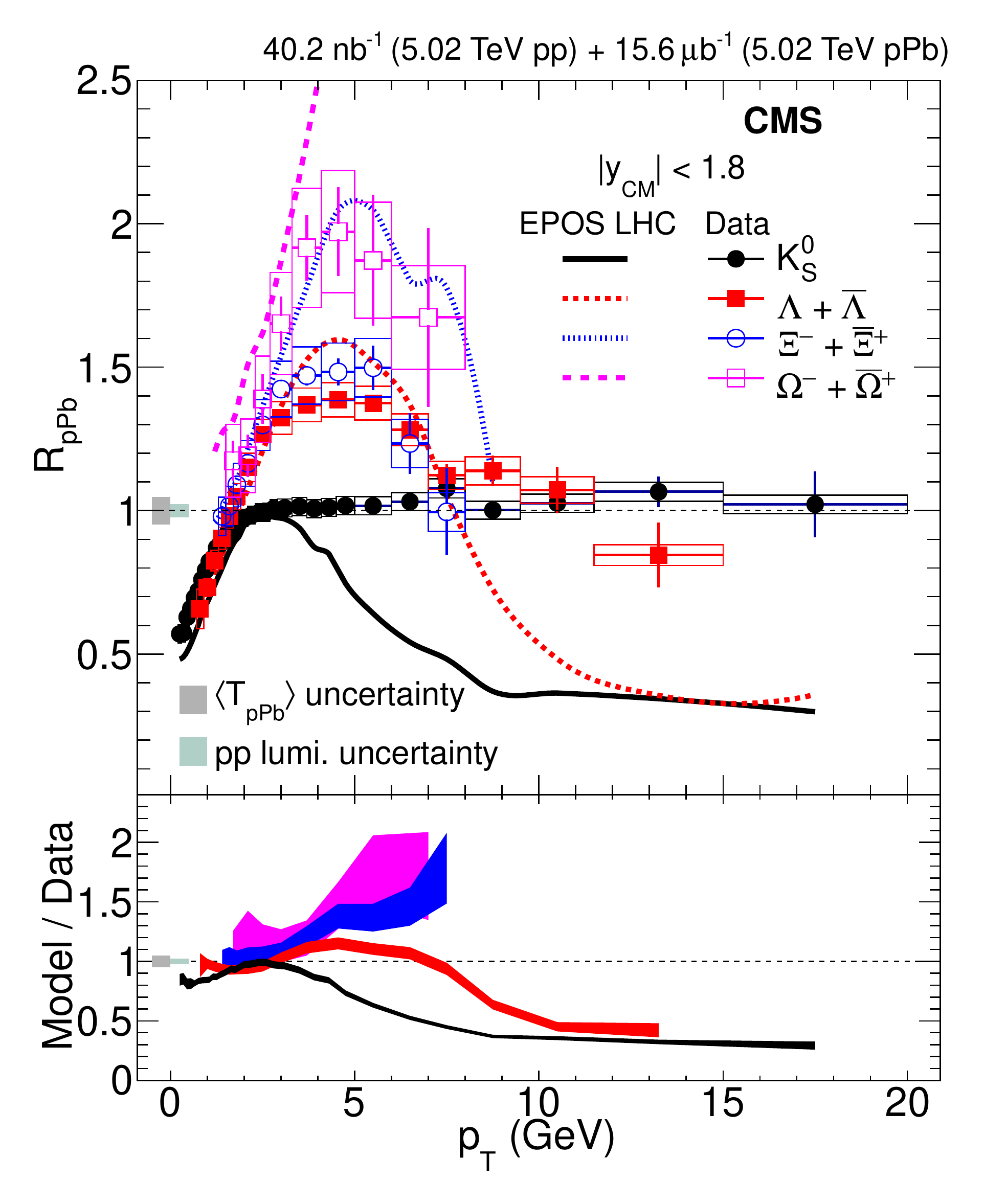}
\caption{
(Upper) Nuclear modification factors for \PKzS\ (black filled circles), \PgLLbar\ (red filled squares), \PgXmXpnew\
(blue open circles), and \PgOmOpnew\ (purple open squares) for $\abs{\ycm} < 1.8$ in \pPb collisions are presented.
The vertical bars correspond to statistical uncertainties, and the horizontal bars represent the bin width, while
the open boxes around the markers denote the systematic uncertainties.
The $\langle \TpPb \rangle$ and \pp\ integrated luminosity
uncertainties are represented by the shaded boxes around unity. The results are compared with the \EPOSLHC predictions,
which include collective flow in \pp\ and \pPb collisions. The data and predictions share the same color for each particle species.
(Lower) The ratios of nuclear modification factors for \PKzS, \PgLLbar, \PgXmXpnew, and \PgOmOpnew\ of the \EPOSLHC\ predictions to the
measurements are shown. The bands represent the combination of statistical and systematic uncertainties.
}
\label{fig:rpa_midy}
\end{figure}

With the efficiency-corrected strange-particle spectra, the \RpPb values of \PKzS, \PgL, \PgXm, and \PgOm\ particles are calculated
in different \ycm ranges. Figure~\ref{fig:rpa_midy} shows the \RpPb of each particle species at $\abs{\ycm} < 1.8$. The \RpPb values
of \PKzS\ are consistent with unity for $\pt > 2\GeV$. For baryons, the \RpPb of both \PgL\ and \PgXm\ reach unity for \pt\ somewhere
between 7 and 8\GeV. This is consistent with the charged-particle \RpPb~\cite{Khachatryan:2015xaa}, which also shows no modification
in the \pt range from 7 to 20\GeV. In the intermediate \pt range from 2 to 7\GeV, an enhancement with clear mass and strangeness-content
ordering is observed for baryons with the greater mass and strangeness corresponding to larger \RpPb. The observed mass ordering is
consistent with expectations from the radial-flow effect in hydrodynamic models~\cite{Pierog:2013ria}. The predictions from \EPOSLHC,
including collective flow in \pp\ and \pPb collisions, are compared with data in Fig.~\ref{fig:rpa_midy}. The calculations indeed predict
clear mass ordering for baryon \RpPb in this \pt\ range, with even stronger mass dependence than observed in data. At higher \pt, \RpPb of
\PKzS\ and \PgL\ calculated from the \EPOSLHC\ model is markedly smaller than the data because of the strong screening in nuclear collisions
in \EPOSLHC. This screening is needed to reduce the number of binary collisions in the initial state in order to produce the correct
multiplicity~\protect{\cite{Pierog:2013ria}}. It is not clear from current measurements whether effects from recombination play a role.
This can be addressed by studies that include identified baryons and mesons with similar masses, such as the measurements of proton
and $\phi$ meson $R_\mathrm{dAu}$ at RHIC~\cite{Adare:2013esx}. To fully understand particle production in this \pt range,
more theoretical calculations including the recombination models are needed. For \pt\ values less than 2\GeV, the predicted \RpPb values
from the \EPOSLHC\ model qualitatively agree with the experimental results for each of the particle species. In this \pt range,  \RpPb for \PKzS\ and \PgL\ become less than unity, as expected for soft particle production.

\begin{figure*}[ht!]
\centering
\includegraphics[width=0.9\textwidth]{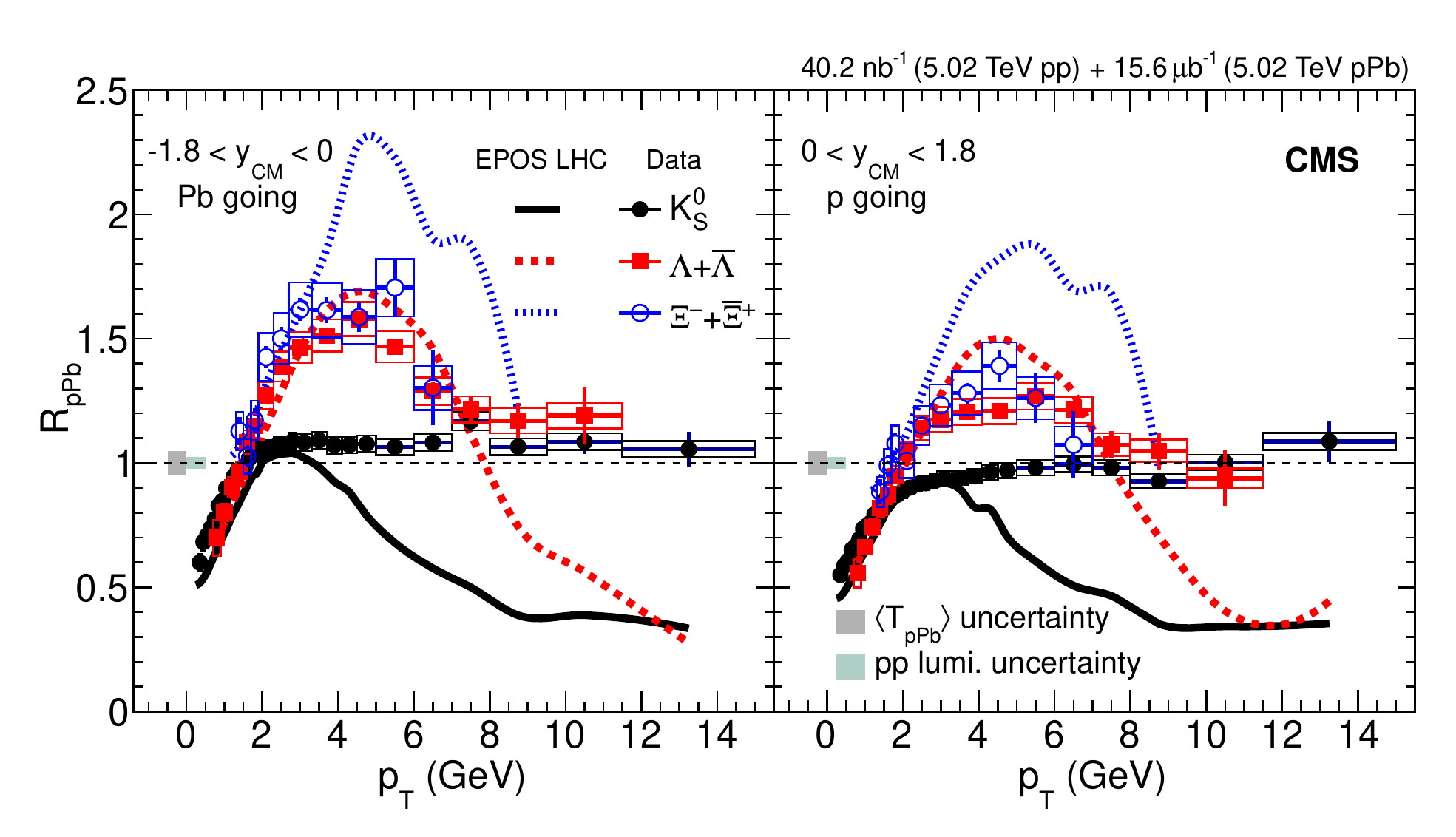}
\caption{Nuclear modification factors of \PKzS\ (black filled circles), \PgLLbar\ (red      filled squares), and \PgXmXpnew\ (blue open circles) particles for $-1.8<\ycm<0$ (Pb going,  left) and $0<\ycm<1.8$ (\Pp going, right) in \pPb collisions are presented. The vertical  bars correspond to statistical uncertainties, and the horizontal bars represent the bin width, while the open boxes around the markers denote the systematic uncertainties. The $\langle \TpPb \rangle$ and \pp\ integrated
luminosity uncertainties are represented by the shaded boxes around unity. The results  are compared with the \EPOSLHC predictions, which include
collective flow in \pp\ and \pPb collisions~\protect{\cite{Pierog:2013ria}}. The data   and predictions share the same color for each particle species.
}
\label{fig:rpa_fory}
\end{figure*}

The \RpPb values of \PKzS, \PgL, and \PgXm\ particles for $-1.8<\ycm<0$ and $0<\ycm<1.8$ are presented as functions of \pt\ in Fig.~\ref{fig:rpa_fory}.
Because of the limitations in the size of the data sample, the \RpPb of the \PgOm\ baryon is not shown in the \Pp- and Pb-going direction separately.
Above $\pt> 2\GeV$,  \RpPb of all three species are found to be larger in the Pb-going direction than the \Pp-going direction, with a stronger splitting between \PKzS\ and baryons in the Pb-going direction.
This trend is consistent with expectations from the radial-flow effect in hydrodynamic models~\cite{Pierog:2013ria, Bozek:2013sda}. The predicted values of \RpPb for \PgXm\ particles from the \EPOSLHC model are larger than those from data in both \Pp-going and Pb-going directions. Momentum broadening from parton multiple scattering as implemented in Ref.~\cite{Wang:2003vy}  predicts a stronger enhancement in the \Pp-going direction, which is inconsistent with the results in Fig.~\ref{fig:rpa_fory}. However, this could be explained by the prediction that this effect is small compared with the nuclear shadowing effect~\cite{Albacete:2013ei} at the LHC energies. The probed parton momentum fraction $x$ in the nucleus is less than 0.02 for the \pt\ and rapidity considered in this analysis. Therefore, these measurements are sensitive to the shadowing effect, and \RpPb should be smaller in the \Pp-going direction because the probed $x$ fractions in the nucleus are smaller. The combined treatment of initial and final-state scatterings described in Ref.~\cite{Kang:2013ufa} is in qualitative agreement with the data.

\begin{figure*}[h!t]
\centering
\includegraphics[width=0.9\textwidth]{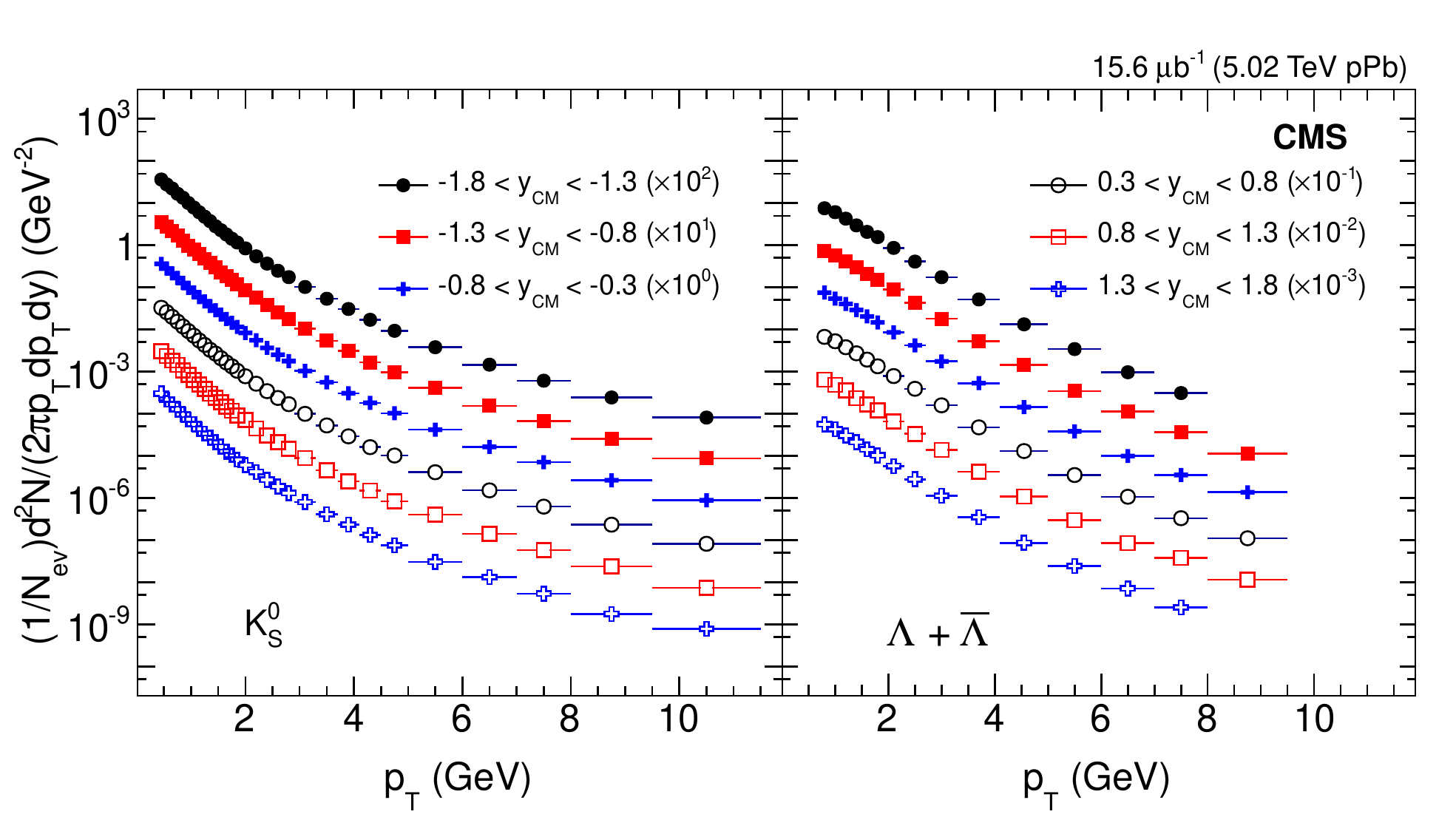}
\caption{The invariant \pt-differential spectra of \PKzS\ (left) and \PgLLbar\ (right)      particles
for $-1.8<\ycm<-1.3$, $-1.3<\ycm<-0.8$, $-0.8<\ycm<-0.3$, $0.3<\ycm<0.8$, $0.8<\ycm<1.3$,
and $1.3<\ycm<1.8$ in \pPb\ collisions at $\sqrtsNN = 5.02\TeV$. Spectra in different   \ycm
ranges are scaled by factors of powers of 10, with $-0.8<\ycm<-0.3$ not scaled.
The vertical bars correspond to statistical uncertainties, which are usually smaller    than the marker size, while the horizontal bars        represent the bin width.
}
\label{fig:spectra_yasym}
\end{figure*}

\subsection{The particle-yield rapidity asymmetry }

\begin{figure}[ht!]
\centering
\includegraphics[width=0.49\textwidth]{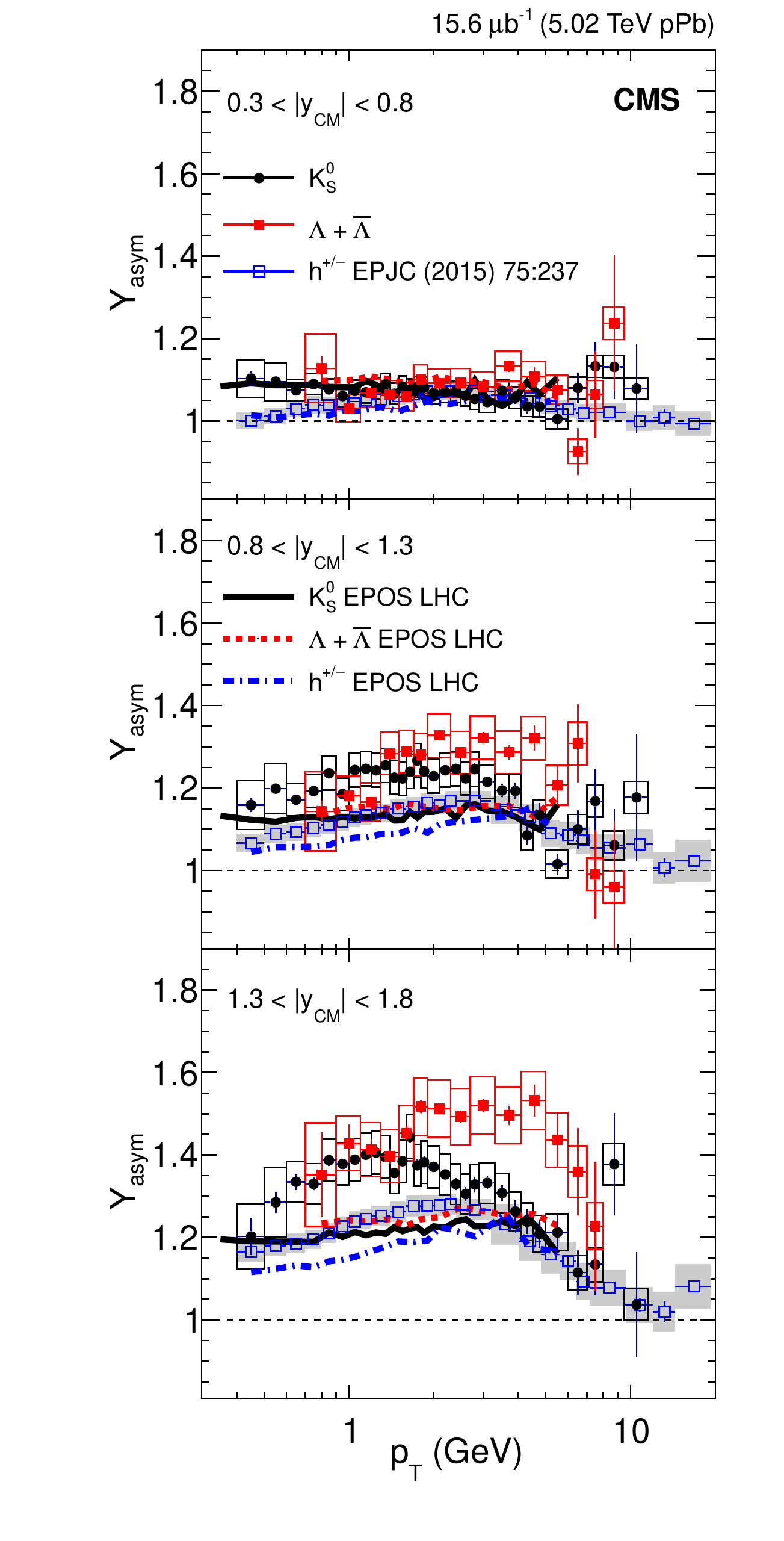}
\caption{The \Yasym of \PKzS\ (black filled circles), \PgLLbar\ (red filled squares), and
charged particles (blue open squares) at $0.3<\abs{\ycm}<0.8$, $0.8<\abs{\ycm}<1.3$,
and $1.3<\abs{\ycm}<1.8$ ($\abs{\eta_\mathrm{CM}}$ ranges for charged particles) in
\pPb collisions at $\sqrtsNN=5.02\TeV$. The vertical bars correspond to statistical
uncertainties, and the horizontal bars represent the bin width, while the boxes around  the
markers denote the systematic uncertainties. The results are compared with the
 \EPOSLHC predictions, which include collective flow in \pp\ and \pPb
 collisions~\protect{\cite{Pierog:2013ria}}.
The data and predictions share the same color for each particle species.
}
  \label{fig:yasym_with12017}
\end{figure}

The invariant \pt-differential spectra of \PKzS\ and \PgL\ for five different \ycm ranges in \pPb collisions at $\sqrtsNN=5.02\TeV$
are presented in Fig.~\ref{fig:spectra_yasym}. Figure~\ref{fig:yasym_with12017} shows the \Yasym (Pb-going direction in the numerator)
as functions of \pt\ for \PKzS, \PgL\ and charged particles~\cite{Khachatryan:2015xaa} for different rapidity (pseudorapidity) ranges.
The observed \Yasym values depend both on \pt and particle species, and these dependencies are more pronounced in the forward
(larger) \ycm ranges. The \Yasym are larger in the forward region, consistent with expectations from nuclear shadowing, and overall
larger than unity in all measured $\abs{\ycm}$ ranges. Significant departures from unity,  and particle species dependencies are seen
away from midrapidity in the region  $1.3<\ycm<1.8$. As a function of \pt for all particle species, the \Yasym values first rise and
then fall, approaching unity at higher \pt. The peak values for \PgL\ are shifted to higher \pt compared with the those of  \PKzS\ and
charged particles, which include a \pt-dependent mixture of charged hadrons. The  \Yasym of \PKzS\ and \PgL\ are larger than those of
charged particles. These detailed structures, with mass dependence and meson-baryon differences, will provide strong constraints on
hydrodynamic and recombination models in which particle species dependencies arise from the differences in mass or number of constituent
quarks, respectively. The results of \Yasym are compared with  
the \EPOSLHC\ predictions 
for \PKzS, \PgL, and inclusive charged particles produced 
in the three \ycm ranges. The \Yasym from \EPOSLHC\
increases from mid-\ycm to forward \ycm, consistent with the trend of the data, but fails to describe the particle-species dependence at
forward \ycm.

\section{Summary}
\label{sec:summary}

The transverse momentum (\pt) spectra of \PKzS mesons, and \PgL, \PgXm, and \PgOm\ baryons (each summed with its antiparticle)  have
been measured in proton-proton and proton-lead collisions in several nucleon-nucleon center-of-mass rapidity (\ycm) ranges. The nuclear modification factors of \PKzS, \PgL, and
\PgXm\ in $\abs{\ycm} < 1.8$, $-1.8<\ycm<0$, and $0<\ycm<1.8$ ranges are measured. In the \pt\ range from 2 to 7\GeV,
enhancements are visible and a clear mass ordering is observed, which is consistent with expectations from radial-flow effects in
hydrodynamic models. For each particle species, the nuclear modification factor \RpPb in the Pb-going side is higher than in the
\Pp-going side. This trend is also consistent with expectations from radial flow.
The rapidity asymmetries \Yasym  in \PKzS\ and \PgL\ yields between
equivalent positive and negative \ycm are presented as functions of \pt\ in $0.3<\abs{\ycm}<0.8$, $0.8<\abs{\ycm}<1.3$, and
$1.3<\abs{\ycm}<1.8$, and compared with those for charged particles. The \Yasym values are larger than unity in all three \ycm ranges with
greater enhancements observed at more forward regions. The mass dependence of \RpPb in the \EPOSLHC model, which includes collective flow, is
stronger than that observed in the data. The model also describes the increasing trend of \Yasym from midrapidity to forward rapidity,
but fails to describe the dependence on particle species at forward rapidity. The results presented in this paper provide new
insights into particle production in \pPb\ collisions at high energies.

\ifthenelse{\boolean{cms@external}}{\clearpage}{}
\begin{acknowledgments}
We congratulate our colleagues in the CERN accelerator departments for the excellent performance of the LHC and thank the technical and administrative staffs at CERN and at other CMS institutes for their contributions to the success of the CMS effort. In addition, we gratefully acknowledge the computing centers and personnel of the Worldwide LHC Computing Grid for delivering so effectively the computing infrastructure essential to our analyses. Finally, we acknowledge the enduring support for the construction and operation of the LHC and the CMS detector provided by the following funding agencies: BMBWF and FWF (Austria); FNRS and FWO (Belgium); CNPq, CAPES, FAPERJ, FAPERGS, and FAPESP (Brazil); MES (Bulgaria); CERN; CAS, MoST, and NSFC (China); COLCIENCIAS (Colombia); MSES and CSF (Croatia); RPF (Cyprus); SENESCYT (Ecuador); MoER, ERC IUT, PUT and ERDF (Estonia); Academy of Finland, MEC, and HIP (Finland); CEA and CNRS/IN2P3 (France); BMBF, DFG, and HGF (Germany); GSRT (Greece); NKFIA (Hungary); DAE and DST (India); IPM (Iran); SFI (Ireland); INFN (Italy); MSIP and NRF (Republic of Korea); MES (Latvia); LAS (Lithuania); MOE and UM (Malaysia); BUAP, CINVESTAV, CONACYT, LNS, SEP, and UASLP-FAI (Mexico); MOS (Montenegro); MBIE (New Zealand); PAEC (Pakistan); MSHE and NSC (Poland); FCT (Portugal); JINR (Dubna); MON, RosAtom, RAS, RFBR, and NRC KI (Russia); MESTD (Serbia); SEIDI, CPAN, PCTI, and FEDER (Spain); MOSTR (Sri Lanka); Swiss Funding Agencies (Switzerland); MST (Taipei); ThEPCenter, IPST, STAR, and NSTDA (Thailand); TUBITAK and TAEK (Turkey); NASU (Ukraine); STFC (United Kingdom); DOE and NSF (USA).

\hyphenation{Rachada-pisek} Individuals have received support from the Marie-Curie program and the European Research Council and Horizon 2020 Grant, contract Nos.\ 675440, 752730, and 765710 (European Union); the Leventis Foundation; the A.P.\ Sloan Foundation; the Alexander von Humboldt Foundation; the Belgian Federal Science Policy Office; the Fonds pour la Formation \`a la Recherche dans l'Industrie et dans l'Agriculture (FRIA-Belgium); the Agentschap voor Innovatie door Wetenschap en Technologie (IWT-Belgium); the F.R.S.-FNRS and FWO (Belgium) under the ``Excellence of Science -- EOS" -- be.h project n.\ 30820817; the Beijing Municipal Science \& Technology Commission, No. Z181100004218003; the Ministry of Education, Youth and Sports (MEYS) of the Czech Republic; the Lend\"ulet (``Momentum") Program and the J\'anos Bolyai Research Scholarship of the Hungarian Academy of Sciences, the New National Excellence Program \'UNKP, the NKFIA research grants 123842, 123959, 124845, 124850, 125105, 128713, 128786, and 129058 (Hungary); the Council of Science and Industrial Research, India; the HOMING PLUS program of the Foundation for Polish Science, cofinanced from European Union, Regional Development Fund, the Mobility Plus program of the Ministry of Science and Higher Education, the National Science Center (Poland), contracts Harmonia 2014/14/M/ST2/00428, Opus 2014/13/B/ST2/02543, 2014/15/B/ST2/03998, and 2015/19/B/ST2/02861, Sonata-bis 2012/07/E/ST2/01406; the National Priorities Research Program by Qatar National Research Fund; the Ministry of Science and Education, grant no. 3.2989.2017 (Russia); the Programa Estatal de Fomento de la Investigaci{\'o}n Cient{\'i}fica y T{\'e}cnica de Excelencia Mar\'{\i}a de Maeztu, grant MDM-2015-0509 and the Programa Severo Ochoa del Principado de Asturias; the Thalis and Aristeia programs cofinanced by EU-ESF and the Greek NSRF; the Rachadapisek Sompot Fund for Postdoctoral Fellowship, Chulalongkorn University and the Chulalongkorn Academic into Its 2nd Century Project Advancement Project (Thailand); the Nvidia Corporation; the Welch Foundation, contract C-1845; and the Weston Havens Foundation (USA).
\end{acknowledgments}

\bibliography{auto_generated}
\cleardoublepage \appendix\section{The CMS Collaboration \label{app:collab}}\begin{sloppypar}\hyphenpenalty=5000\widowpenalty=500\clubpenalty=5000\input{HIN-16-013-authorlist.tex}\end{sloppypar}
\end{document}

%% file: HIN-16-013-authorlist.tex
\vskip\cmsinstskip
\textbf{Yerevan Physics Institute, Yerevan, Armenia}\\*[0pt]
A.M.~Sirunyan, A.~Tumasyan
\vskip\cmsinstskip
\textbf{Institut f\"{u}r Hochenergiephysik, Wien, Austria}\\*[0pt]
W.~Adam, F.~Ambrogi, E.~Asilar, T.~Bergauer, J.~Brandstetter, M.~Dragicevic, J.~Er\"{o}, A.~Escalante~Del~Valle, M.~Flechl, R.~Fr\"{u}hwirth\cmsAuthorMark{1}, V.M.~Ghete, J.~Hrubec, M.~Jeitler\cmsAuthorMark{1}, N.~Krammer, I.~Kr\"{a}tschmer, D.~Liko, T.~Madlener, I.~Mikulec, N.~Rad, H.~Rohringer, J.~Schieck\cmsAuthorMark{1}, R.~Sch\"{o}fbeck, M.~Spanring, D.~Spitzbart, A.~Taurok, W.~Waltenberger, J.~Wittmann, C.-E.~Wulz\cmsAuthorMark{1}, M.~Zarucki
% ...
\vskip\cmsinstskip
% ...
1:  Also at Vienna University of Technology, Vienna, Austria\\
% ...
